\documentclass[twocolumn,amsmath]{revtex4}

\usepackage{amsfonts}
\usepackage{amsmath}
\usepackage{amssymb}
\usepackage{graphicx}%

\begin{document}

\title{Impact of electron shell excitations on the energy spectrum \\of $\beta$-electrons in neutrinoless double-$\beta$ decay}
\author{M. I. Krivoruchenko, K. S. Tyrin}
\affiliation{National Research Centre ''Kurchatov Institute'', Pl. Akademika Kurchatova 1, 123182 Moscow, Russia}
\author{F. F. Karpeshin}
\address{D. I. Mendeleev All-Russian Research Institute of Metrology (VNIIM) \\
Moskovsky Ave. 19$\mathrm{,}$ 190005 Saint Petersburg}

\begin{abstract}
The electron shell of the daughter atoms often appears excited in the double-$\beta$ decays,
which causes a change in the energy taken away by $\beta$-electrons. The average value and variance
of the excitation energy of the electron shell of the daughter atom are calculated for the double-$\beta$
decay of germanium $_{32}^{76}\mathrm{Ge} \rightarrow _{34}^{76}\mathrm{Se}^*+2\beta^-(+~2\bar{\nu_e})$
in both the Thomas--Fermi model and the relativistic Dirac--Hartree--Fock theory.
Using the results obtained, a two-parameter model of the energy spectrum of $\beta$-electrons
in the neutrinoless mode is constructed, taking into account energy redistribution in the decay channels.
The shift in total energy of $\beta$-electrons is found to be under
50 eV at a confidence level of 90\%
.
The average excitation energy, on the other hand, is an order of magnitude higher and equal to
$\sim 400$ eV, while the square root of the variance is equal to $\sim 2900$ eV,
which is presumably explained by the contribution of the core electrons to the
energy characteristics of the process. The probability is nearly saturated with excitations
with a small amount of released energy, which is common for the outermost electrons.
The distortion of the peak shape of the neutrinoless double-$\beta$ decay should be taken into consideration when analyzing
data from detectors with a resolution of $\sim 100$ eV or higher.
\end{abstract}

\maketitle

Neutrinos are one of the most promising tools for investigating new physics beyond the Standard Model.
The study of $\beta$-processes sensitive to neutrino mixing, mass, and nature (Dirac/Majorana) presents
an attractive approach in the quest for extensions of the Standard Model.

Experimental observation of neutrinoless double-$\beta$ decay
would significantly narrow the range of possibilities for the generalization.
Such an observation would prove the existence of a Majorana neutrino mass, as well as a violation of the
conservation of the total lepton number (for a review see, e.g., \cite {Bilenky:2018}).
The neutrinoless double-$\beta$ decay is registered by a narrow peak at the kinetic energy of $\beta$-electrons, $T$,
equal to the decay energy, $Q$.
Channels with the electron shell of the daughter atom in the ground state do not always dominate in $\beta$
processes \cite{Bambynek:1977,Krivoruchenko:2020}, which leads, in particular, to an increased rate of
neutrinoless double-electron capture in non-resonant modes \cite{Karpeshin:2020,Karpeshin:2022,Karpeshin:2023}.
Excitation of electron shells and chemical shifts of atomic levels, which depend on the aggregate state of the substance \cite{Lindgren:2004},
affect the energy spectrum in the neutrinoless double-$\beta$ decay.
The energy peak at $T=Q$ is used as an indicator of the neutrinoless double-$\beta$ decay, and the search for it is being carried out
by the collaborations CUORE \cite{Alfonso:2015}, EXO \cite{Anton:2019}, and KamLAND-Zen \cite{Gando:2016}.
The SuperNEMO \cite{Arnold:2015} collaboration can observe multielectron modes.
The GERDA collaboration \cite{GERDA:2017} determined an upper limit on the half-life $_{32}^{76}\mathrm{Ge} \rightarrow _{34}^{76}\mathrm{Se}^*+ 2\beta^-$ of $5.3\times10^{25}$ years with a confidence level of 90\%.
To interpret the results in terms of the mass of the Majorana neutrino, the uncertainties in the axial coupling
constants of nucleons and the nuclear matrix elements should be controlled \cite{Suhonen:2017,Simkovic:2008}.
In this paper, the effect of electron shell excitations on the energy spectrum of $\beta$-electrons
in the double-$\beta$ decays is investigated.

The nucleus of a $\beta$-decaying atom experiences a charge change of $\Delta Z = \pm1, \pm2$. The change in charge
acts on the electron shell as a sudden perturbation, causing the electrons of the daughter atom to move with
an appropriate probability to higher discrete levels (shake-up) or to the continuous spectrum
(shake-off).
The energy spectrum of $\beta$-electrons is determined by
a decay channel, where all electrons maintain their initial quantum numbers,
and decay channels with electron shell excitation.
In the second case, the effective reaction energy decreases, modifying the energy spectrum of $\beta$-electrons.

To calculate the energy spectrum, it is necessary to sum over all the multiparticle channels with their individual amplitudes,
taking into account the modified reaction energy.
In a configuration corresponding to a specific excitation of the shell, the wave function of the parent-atom
electrons is projected onto the wave function of the daughter-atom electrons. As a rule, each such amplitude
needs a complex multiparticle numerical simulation. The Coulomb problem is also unique in that there are
an infinite number of decay channels.

The computation of the average excitation energy of the daughter atomic shell and its variance,
however, is quite straightforward within the Thomas--Fermi model and the relativistic Dirac--Hartree--Fock
(DHF) formalism.
The results obtained with these methods can be used to construct a simple two-parameter
probability distributions
of the energy taken away by $\beta$-electrons.

The non-relativistic Hamiltonian for an atom with a nuclear charge of $Z$ and $N$ electrons has the form
\begin{equation} \label{hamiltonian}
\hat{H}_{Z,N}=\sum_{i=1}^{N}\left( \frac{1}{2}\mathbf{p}_{i}^{2} - \frac{Z}{|\mathbf{r}_{i}|} \right)
+ \sum_{i<j}^{N}\frac{1}{|\mathbf{r}_{i}-
\mathbf{r}_{j}|},
\end{equation}
where $\mathbf{p}_{i}$ are the momenta and $\mathbf{r}_{i}$ are the coordinates of the electrons;
hereafter, the system of atomic units is used: $\hbar = m = e = 1$, $c = 137$,
where $m$ is the mass of the electron, $e$ is the charge of the proton, and $c$ is the speed of light.
The ground state is denoted by $|Z,N\rangle$.
The ground-state binding energy of the electrons, $E_{Z,N}$, is the eigenvalue of $\hat{H}_{Z,N}$.

In neutrinoless double-$\beta$ decay, the nuclear charge increases by two units, and the parent-atom ground state becomes
a superposition of energy eigenstates of the daughter-ion Hamiltonian:
\begin{equation} \label{decompose}
\hat{H}_{Z+2,Z} = \hat{H}_{Z,Z} -2 \sum_{i=1}^{Z}\frac{1}{{r}_{i}},
\end{equation}
where $r_i = |\mathbf{r}_{i}|$.
The second term acts as a sudden perturbation.
In the daughter ion, the average energy of the $Z$ electrons that make up the parent atomic shell is equal to
\[
\langle Z,Z|\hat{H}_{Z+2,Z}|Z,Z\rangle = E_{Z,Z} + 2Z^{-1} E_{Z,Z}^{\mathrm{C}}, 
\]
where $E_{Z,Z}^{\mathrm{C}}$  is the Coulomb interaction energy of $Z$ electrons with the nucleus:
\begin{eqnarray} \label{Coulomb}
E_{Z,Z}^{\mathrm{C}} = - Z \sum_{i=1}^{Z} \langle Z,Z| \frac{1}{r_i}|Z,Z\rangle .
\end{eqnarray}
The average excitation energy of the electron shell becomes
\begin{eqnarray} \label{C}
\mathcal{M} = E_{Z,Z} + 2 Z^{-1} E_{Z,Z}^{\mathrm{C}} - E_{Z+2,Z}.
\end{eqnarray}
The binding energies of neutral atoms $E_{Z,Z}$ are tabulated in Refs.~\cite{Lu:1971, Desclaux:1973, Clementi:1974, HUANG:1976}
and can be computed with the use of computer programs that determine the electron shell structures of atoms
(see, e.g., \cite{Dyall:1989}). The value of $E_{Z+2,Z}$ differs from the total binding energy of electrons in the neutral atom $E_{Z+2,Z+2}$
by the ionization energe of two valence electrons, which does not exceed $\sim 20$ eV.
The values of the Coulomb energy $E_{Z,Z}^{\mathrm{C}}$ obtained with the use of the DHF method
are reported in Ref.~\cite{HUANG:1976}.


The variance of excitation energy reads
\begin{eqnarray}
\mathcal{D} = \langle Z,Z|\hat{H}_{Z+2}^2|Z,Z\rangle - \langle Z,Z|\hat{H}_{Z+2}|Z,Z\rangle^2. \label{disp}
\end{eqnarray}
Given Eq. (\ref{decompose}) and the fact that $E_{Z,N}$ is the eigenvalue of $\hat{H}_{Z,N}$, we obtain
\begin{eqnarray*} \label{211}
\frac{1}{4} \mathcal{D} = \sum_{i=1}^Z\sum_{j=1}^Z\langle Z,Z|\frac{1}{r_i} \frac{1}{r_j} |Z,Z\rangle -
\langle Z,Z| \sum_{i=1}^Z \frac{1}{r_i} |Z,Z\rangle ^2. 
\end{eqnarray*}
In the first term, we neglect the exchange contributions and factorize the matrix elements for $i\neq j$:
\[
\langle Z,Z| \frac{1}{r_ir_j}|Z,Z\rangle \approx \langle Z,Z| \frac{1}{r_i}|Z,Z\rangle
\langle Z,Z| \frac{1}{r_j} |Z,Z\rangle
.\]
The cancellation of non-diagonal terms caused by the factorization allows to give the variance in the form
\begin{eqnarray} \label{dispdir}
\frac{1}{4} \mathcal{D} \approx \sum_{i=1}^Z \langle Z,Z| \frac{1}{ r_i^2}|Z,Z\rangle - \sum_{i=1}^Z \langle Z,Z| \frac{1}{r_i}|Z,Z\rangle^2.
\end{eqnarray}
The electron wave functions of the Roothaan-Hartree-Fock formalism \cite{Clementi:1974}
can be used to estimate the exchange effects and demonstrate that the exchange contribution to the variance is under 10\%.

Equations (\ref{decompose}) -- (\ref{dispdir}) are also valid in the relativistic framework
using the Dirac-Coulomb Hamiltonian as a zero-order approximation and the Breit potential
to describe the electron-electron interaction in the order of $1/c^2$ \cite{Grant:2008}.


The total binding energy and Coulomb interaction energy in the Thomas--Fermi model
are given by the equations $E_{Z,Z} = -0.764 Z^{7/3}$ and $E_{Z,Z}^{\mathrm{C}} = 7E_{Z,Z}/3$, respectively  (see, e.g., \cite{LL:1987}).
The virial theorem establishes the relationship between $E_{Z,Z}$ and $E_{Z,Z}^{\mathrm{C}}$.
The average excitation energy is equal to
\begin{equation}
\mathcal{M} = \left(1+\frac{14}{3Z}\right)E_{Z,Z} - E_{Z+2,Z}.
\end{equation}
In the double-$\beta$ decay of germanium,
each of the two missing 4$p$ electrons of selenium is bound at 9.752 eV, while the sum of the first two
selenium ionization energies is 30.948 eV \cite{Kramida:2022}.
Taking into account the correction, we find $\mathcal{M} = 382$ eV.

The electron density is expressed in terms of the wave function of atomic shells according to
\begin{equation} \label{216}
n(\mathbf{r}) = \sum_{i=1}^Z \langle Z,Z| \delta(\mathbf{r} - \mathbf{r}_{i})|Z,Z\rangle.
\end{equation}
The first term in Eq.~(\ref{dispdir}) can obviously be reduced to the form
$\int d\mathbf{r} r^{-2} n(\mathbf{r})$.
The parametrization of shielding function for neutral atoms of Ref.~\cite{Mason:1964} can be
utilized to calculate the integral.
The integral diverges as $\sim dr/r^{3/2}$ at short distances. The semi-classical
approximation has a distance restriction $1/Z \lesssim r$. The electron density is taken to
be constant and equal to $n(r=1/Z)$ for numerical calculations in the range $0 \leq r \leq 1/Z$.

To evaluate the second term in Eq.~(\ref{dispdir}), we apply the well-known inequality,
according to which the arithmetic mean does not exceed the quadratic mean:
\begin{eqnarray}
\frac{1}{Z}\sum_{i=1}^Z \langle Z,Z| \frac{1}{{r}_i}|Z,Z\rangle^2 \geq
\left( \frac{1}{Z}\sum_{i=1}^Z \langle Z,Z| \frac{1}{{r}_i}|Z,Z\rangle \right)^2 \nonumber \\
= \left(\frac{1}{Z}\int d\mathbf{r} \frac{1}{r} n(\mathbf{r}) \right)^2.~~~~~~~~~~~~~~~~~~ \label{ineq}
\end{eqnarray}
The right side equals $(Z^{-2}E_{Z,Z}^{\mathrm{C}})^2$. The inequality determines
the upper bound of the variance:
\begin{equation} \label{ineq}
\frac{1}{4} \mathcal{D} \leq \frac{1}{4} \bar{\mathcal{D}} = \int d\mathbf{r} \frac{1}{{r}^2} n(\mathbf{r}) - Z^{-1} \left( \int d\mathbf{r} \frac{1}{{r}} n(\mathbf{r}) \right)^2.
\end{equation}
The estimated value for germanium double-$\beta$ decay is $\bar{\mathcal{D}}^{1/2} $ = $ 2160$ eV.

The first negative moment in the germanium atom has a value of $\langle 1/r\rangle = 4.99$ 
\cite{Lu:1971}.
The relativistic DHF approach is also implemented in the RAINE software package \cite{RAINE,Band:1986},
which calculates the position of the selenium ground-state energy with respect to the germanium ground-state energy.
The average excitation energy $\mathcal{M} = 300$ eV is found by substituting the values obtained
into Eq.~(\ref{dispdir}), which is qualitatively compatible with the Thomas-Fermi result.
The values of $E_{Z,Z}$ and $E_{Z,Z}^{\mathrm{C}}$ of Huang et al. \cite{HUANG:1976}
result in $\mathcal{M} = 400$ eV.

The variance (\ref{dispdir}) is calculated in terms of the first and second negative moments
of the electron distribution in the germanium atom.
Using the values of the second moments of the electron distribution in atoms given by Desclaux \cite{Desclaux:1973}
one finds $\mathcal{D}^{1/2} = 2870$ eV.
A qualitative consistency is also seen when the Thomas--Fermi value of $2160$ eV is compared to the obtained value.


The values of $\mathcal{M}$ and $\mathcal{D}$ are substantially greater than expected for channels associated with valence shell excitations.
The shake-up contribution of the outer shells to $\mathcal{M}$ and $\mathcal{D}$ does not exceed $\sim 10$ eV.
The binding energy determines the convergence of the integral over a continuous energy spectrum,
so the shake-off contribution to $\mathcal{M}$ and $\mathcal{D}$ is comparable to the shake-up contribution.
At the same time, valence shell excitations dominate the probability, because a change in the nuclear charge $\Delta Z = 2$
results in a considerable, in relative units, change in the screened potential at the atomic border.

Due to their lower energy, the core electrons have a small probability of being excited. They may considerably
contribute, however, to both the average excitation energy and the variance.
The change in the nuclear charge $\Delta Z$ is much less than $Z$, allowing perturbation theory
to be used to find the probability of the K--electron shake-off \cite{Feinberg:1941,Migdal:1941} (see also \cite{LL:1987}).
In the leading order, the probability, the average excitation energy, and the average square of excitation energy are equal
$\Delta\mathcal{P}= 0.65\Delta Z^2 /Z^2$, $\Delta\mathcal{M}= 0.66\Delta Z^2$ and $\Delta\mathcal{M}_2 = 1.87 \Delta Z^2 Z^2$, respectively.
In the double-$\beta$ decay of germanium, $\Delta\mathcal{P} = 2.6\times 10^{-3}$, $\Delta\mathcal{M}= 72$ eV and $\Delta\mathcal{D}\equiv\Delta\mathcal{M}_2 - \Delta \mathcal{M}^2 = (2380$~eV$)^2$. $\Delta\mathcal{M}$ and $\Delta\mathcal{D}$ are comparable to $\mathcal{M}$ and $\mathcal{D}$.
The average and variance of the excitation energy are therefore strongly influenced by the K-electron shake-off.

\begin{figure}[t]
\vspace{- 49 mm}
\begin{center}
\includegraphics[width=0.95\columnwidth]{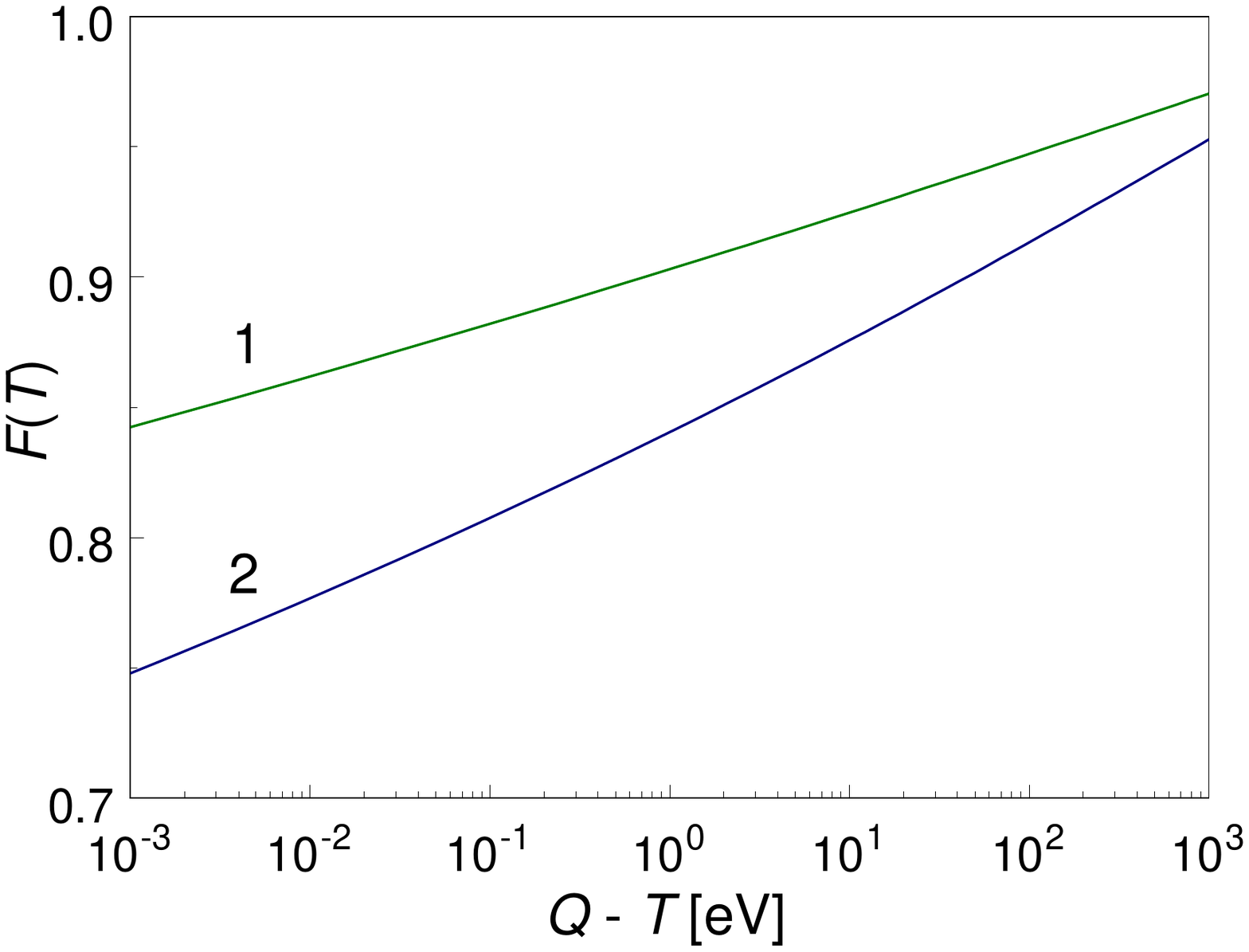}
     \caption{\footnotesize{
      The probability distribution function of the kinetic energy of $\beta$-electrons, $T$, taking into account the overlap factor, $K_Z$, of the electron shells of atoms according to Eq.~(\ref{PDF}). Lines 1 and 2 correspond to the average excitation energy of the electron shell, $\mathcal{M} = 300$ eV and $400$ eV, respectively, and the square root of the variance, $\mathcal{D}^{1/2} = 2870$ eV. Numerical values characterize germanium double-$\beta$ decay, with $Q=2039.061(7)$ keV being the reaction energy \cite{GERDA:2017}.
      }  \label{fig1}}
      \end{center}
      \end{figure}

In $\beta$ processes, the parent-atom electron shell evolves into the daughter-ion electron shell in the ground state with a finite probability
that is not equal to unity \cite{Bambynek:1977,Krivoruchenko:2020,Ge:2021}.
To determine the overlap amplitude $K_Z = \langle \mathrm{Se~III}|\mathrm{Ge}\rangle $ in germanium double-$\beta$ decay,
the electron wave functions of the selenium ion $|\mathrm{Se~III}| \rangle $ in the ground state and the electron wave functions of the germanium atom $|\mathrm{Ge} \rangle$ in the ground state are constructed using the G\textsc{rasp}-2018 software package, which implements the relativistic DHF method
\cite{Dyall:1989,Grant:2008}.
The value of $K_Z$ is sensitive
to the approximations used due to the noticeable difference in binding energies of valence electrons
with the same quantum numbers in $|\mathrm{Ge} \rangle$ and $|\mathrm{Se~III}| \rangle $. Calculations based on G\textsc{rasp}-2018 result in $K_Z = 0.575$.

In the neighborhood of $T=Q$, the kinetic energy of $\beta$-electrons with account of the energy loss
for the excitation and ionization of the daughter atom
is described by the probability distribution function
\begin{eqnarray}
dF(T)&=&\left(K^{2}_{Z}\delta(T - Q) \right. \nonumber \\
&+& \left. (1-K^{2}_{Z})w(1 - T/Q)/Q)\right)dT,
\end{eqnarray}
where $K^{2}_{Z}$ is the probability of the electron shell of the parent atom to evolve
into the ground state of the daughter ion, and
$w(x)$ is the probability density function
of the excitation energy $\epsilon = Q - T$, measured in units of $Q$.
The second term leads to a smearing of the peak.

The average value, the variance, and the overlap amplitude, $K_Z$, can be used
to construct a simple two-parameter model for the energy spectrum of
$\beta$-electrons. A wide range of random processes
is well described by the beta distribution \cite{Korolyuk:1985}
\begin{equation}
\label{wb}
w(x)=\frac{\Gamma(\alpha+\beta)}{\Gamma(\alpha)\Gamma(\beta)}x^{\alpha-1}(1-x)^{\beta-1}
\end{equation}
for a random variable $x$ taking values on the interval $(0,1)$;
$\alpha, \beta > 0$ are free parameters.
In the case under consideration, $x = \epsilon/Q$.
The averages of $x$ and $x^2$ equal
$m={\alpha}/(\alpha+\beta)$ and $m_2=\alpha(\alpha + 1)/((\alpha+\beta)(\alpha+\beta+1))$, respectively.
The equations $(1-K^{2}_{Z})m=\mathcal{M}/Q$ and $(1-K^{2}_{Z})m_2=(\mathcal{D} + \mathcal{M}^2)/Q^2$
allow to determine the parameters $\alpha$ and $\beta$ in terms of $\mathcal{M}$, $\mathcal{D}$, and $K_Z$.
As an example, let us consider the germanium-76 isotope used in the GERDA collaboration experiments.
For $\mathcal{M} = 300$ eV and $400$ eV and the variance $\mathcal{D}^{1/2}= 2870$ eV
we find
$\alpha = 0.016$, $\beta = 74$ and
$\alpha = 0.029$, $\beta = 99$, respectively.
The two-parameter gamma distribution \cite{Korolyuk:1985}, describing random quantities on the half-axis $(0,+\infty)$,
is also suitable for the modeling, since the condition $Q \gg \mathcal{M}, \mathcal{D}^{1/2}$
allows to extend the integration by $x = \epsilon/Q$ to the half-axis $(0,+\infty)$.
In the physically interesting region $T \sim Q$, the beta and gamma distributions practically coincide
because of $\beta\gg 1$.

The probability density is singular, but integrable at $T=Q$. Figure \ref{fig1} shows
the distribution function
\begin{equation} \label{PDF}
F(T) = K_Z^2 + (1 - K_Z^2)\int_{T}^{Q}w(1 - T^{\prime}/Q)dT^{\prime}/Q,
\end{equation}
which determines the probability
of $\beta$-electrons to have an energy that differs from $Q$ by no more than $Q - T >0$.
For $\mathcal{M} = 300$ eV
with a probability of 90\%
, the energy of $\beta$-electrons deviates from $Q$ by no more than 1 eV.
For $\mathcal{M} = 400$ eV with the probability of 90\%,
the deviation is less than 50 eV.
The probability increases with $Q - T$  roughly logarithmically, i.e., quite slowly.
The modern detectors measure the energy released in double-$\beta$ decay with a resolution of $\sim 1$ keV,
which significantly complicates the observation of effects associated with the excitation of electron shells.
The processes of shaking K--electrons are accompanied by the energy release of $\sim 10$ keV, however, these processes are rare.

To summarize, the electron shell of the daughter atom in double-$\beta$ decay is excited with high probability.
In $\beta$ processes, transitions of the valence electrons to discrete states or to a continuous
spectrum dominate the probability.
The average excitation energy of a daughter atom's electron shell,
$\mathcal{M}$, and its variance, $\mathcal{D}$, are estimated for germanium $0\nu 2\beta$ decay
using the Thomas--Fermi model and the relativistic Dirac--Hartree--Fock formalism.
The reported values $\mathcal{M}\sim 400$ eV and $\mathcal{D}^{1/2}\sim 2900$ eV greatly exceed the
values typical for valence electron processes.
The analysis of the energy characteristics reveals
that less probable K--electron shake-off processes contribute significantly to the
average excitation energy and its variance.
The distortion of the peak shape in the neutrinoless double-$\beta$ decay caused by electron shell excitations
becomes important for the interpretation of observational data when the resolution of detectors increases to $\sim 100$ eV.

The work was supported by the Russian Science Foundation under grant No. 23-22-00307.

\end{document}